\documentclass[useAMS,usenatbib,usegraphicx]{mn2e}

\usepackage[british]{babel}
\usepackage{times}
\usepackage{latexsym,amssymb}
\usepackage{journalnames}

\newcommand{\na}{Na\textsc{I}}

\def\kms{$\mbox{km s}^{-1}$}

\def\co{$^{12}$CO\,(2--0)}
\def\dco{$D_{CO}$}

\def\rb{$r_\mathrm{b}$}

\title[The complex nature of the nuclear star cluster in FCC\,277]{The complex nature of the nuclear star cluster in FCC\,277\thanks{Based on observation collected at the ESO Paranal La Silla Observatory, Chile, Prog. ID 380.B-0530, PI L. Infante}
}
 \author[M. Lyubenova et al.]{%
 Mariya Lyubenova$^1$\thanks{e-mail: lyubenova@mpia.de}, 
Remco C.~E. van den Bosch$^1$, 
Patrick C\^{o}t\'{e}$^2$, 
Harald Kuntschner$^3$, 
\newauthor
Glenn van de Ven$^1$, 
Laura Ferrarese$^2$, 
Andr\'{e}s Jord\'{a}n$^{4,5}$, 
Leopoldo Infante$^4$,
Eric W. Peng$^{6,7}$
\\
$^1$Max Planck Institute for Astronomy, K\"onigstuhl 17, D-69117 Heidelberg, Germany\\
$^2$National Research Council of Canada, Victoria, BC, V9E 2E7, Canada\\
$^3$ESO, Karl-Schwarzschild-Str. 2, D-85748 Garching bei  M\"unchen, Germany\\
$^4$Departamento de Astronom\'ia y Astrof\'isica, Pontificia Universidad Cat\'olica de Chile, Av. Vicu\~na Mackenna 4860, Macul 7820436, Santiago, Chile\\
$^5$The Milky Way Millennium Nucleus, Av. Vicu\~{n}a Mackenna 4860, 7820436 Macul, Santiago, Chile\\
$^6$Peking University, Department of Astronomy, 5 Yiheyuan Road, Haidian, Beijing 100871, China\\
$^7$Kavli Institute for Astronomy and Astrophysics, Peking University, 5 Yiheyuan Road, Haidian, Beijing 100871, China
}

\pagerange{\pageref{firstpage}--\pageref{lastpage}} \pubyear{2013}

\begin{document}

\label{firstpage}

\maketitle

\begin{abstract}
Recent observations have shown that compact nuclear star clusters (NSCs) are present in up to 80\% of galaxies. However, detailed studies of their dynamical and chemical properties are confined mainly to  spiral galaxy hosts, where they are more easily observed. In this paper we present our study of the NSC in FCC\,277, a nucleated elliptical galaxy in the Fornax cluster. We use a combination of adaptive optics assisted near-infrared integral field spectroscopy, Hubble Space Telescope imaging, and literature long slit data. We show that   while the NSC does not appear to rotate within our detection limit of $\sim$6\,\kms\/, rotation is detected at larger radii,  where the isophotes appear to be disky, suggesting the presence of a nuclear disk. We also observe a  distinct central velocity dispersion drop that is indicative  of a dynamically cold rotating sub-system. Following the results of orbit-based dynamical modelling, co-rotating as well as counter-rotating stellar orbits are simultaneously needed to reproduce the observed kinematics. We find evidence for varying stellar populations, with the NSC and nuclear disk hosting younger and more metal rich stars than the main body of the galaxy.  We argue that  gas dissipation and some level of merging have likely played an important role in the formation of the nucleus of this intermediate-mass galaxy.  This is in contrast to NSCs  in low-mass early-type galaxies, which may have been formed primarily through the infall of star clusters.
\end{abstract}


\section{Introduction}
\label{sec:intro}

It is  now believed that up to 80\%  of all galaxies host nuclear star clusters (NSCs) in their centres \citep[e.g.][]{carollo98,boeker02,cote06}. Typically, low- and intermediate-luminosity galaxies show a central light excess inside a characteristic radius \rb$\sim0.02 R_{eff}$ above the  inner extrapolation of the global light profile \citep{cote07}. These NSCs usually reside in the photometric centre of the galaxy \citep{binggeli00,boeker02} and their location overlaps with the kinematic centre \citep{neumayer11}. NSCs are usually brighter than typical globular clusters, compact ($r\sim5$ pc), massive ($M\sim10^{7}M_{\odot}$),  may be flattened, and often contain multiple stellar populations  and complex structures \citep[e.g.][]{walcher05, walcher06,cote06,rossa06,seth06,seth08,barth09,turner12,pl12}. Their masses seem to correlate with the mass of the host galaxies \citep{ferrarese06,wh06}, extending the super-massive black holes scaling relations to the low-mass end of galaxies.  In some cases, NSCs  appear to co-exist with central black holes \citep[e.g. review by][and references therein]{graham09} and recently \citet{neumayer12} suggested that NSCs may be the precursors of massive black holes in galaxy nuclei. However, only  handful of detailed studies on the properties of NSCs exists and these are mainly focused on such objects in late-type galaxies. Characterising NSCs in early-type galaxies is a non trivial task, both because of the high surface brightness of the underlying galaxy, and because of  the NSCs compact sizes.

With the availability of adaptive optics fed integral field unit (IFU) instruments, this task is  now becoming feasible. For example, \citet{seth10} have shown that the nucleus of NGC\,404, a nearby S0 galaxy,   hosts several morphologically and dynamically distinct components. The NSC in this galaxy shows a modest rotation aligned with the galaxy, a  gas disk that rotates perpendicularly to the stars, and probably an intermediate-mass black hole ($\sim10^{5}\,M_{\odot}$). Such complicated structure inevitably poses the question of how NSCs have formed. Currently, there are two main scenarios proposed. The first involves the dissipationless infall of star clusters to the galaxy centre due to dynamical friction \citep{tremaine75}. The second suggests NSCs to be the result of dissipational sinking of gas to the galactic centre \citep{mihos94}.

%
\begin{figure}
\resizebox{\hsize}{!}{\includegraphics{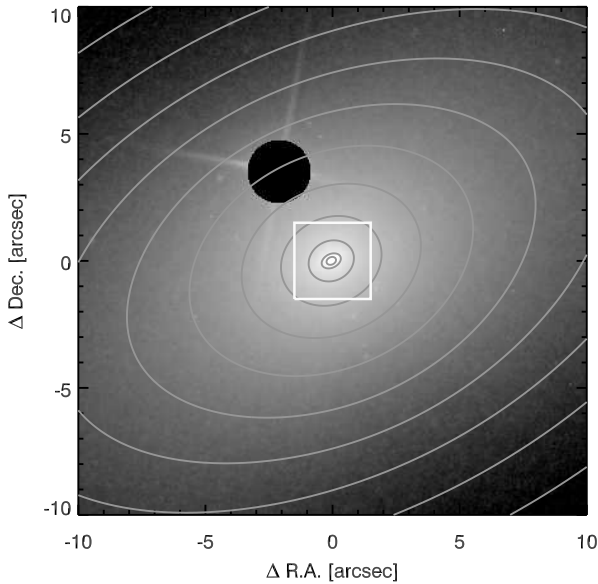}}
\caption{\label{fig:acs_mge} HST/ACS $z$-band image of FCC\,277. The star that we used as a NGS for the AO correction is masked out here. The SINFONI FoV ($3\arcsec\times3\arcsec$) is indicated with a white box. The overlaid contours are from the MGE light model,  discussed in Sect.~\ref{sec:dyn_model}.
}
\end{figure}

Numerical simulations of globular clusters infall have had certain success in reproducing the observed surface brightness profiles of nucleated galaxies , although with larger nuclei sizes comparing to  what is observed \citep[e.g.][]{oh00,cd08a,cd08b}. However, more recently \citet{hartmann11} showed that star cluster accretion onto a pre-existing nuclear disk did not produce the observed line-of-sight kinematics of NSCs. They suggested that purely stellar dynamical mergers cannot  be solely responsible for the formation of NSCs and that gas dissipation  must also play a significant role in assembling the cluster's mass. What is the exact origin of this gas and how it gets transported to the galaxy nucleus is still under debate. \citet{bekki06} have shown that the dissipative merging of stellar and gaseous clumps formed from nuclear gaseous spiral arms in a gas disk eventually produce nuclei that rotate, are flattened and have a range of ages and metallicities. \citet{pflamm09} concluded that compact star clusters with masses $\geq 10^{6} M_{\odot}$ act as cloud condensation nuclei and are able to accrete gas recurrently from a warm interstellar medium. This may cause further star formation events and account for multiple stellar populations in the most massive globular and nuclear star clusters. Recently, \citet{turner12} concluded that the dominant mechanism for nucleus growth in low mass  early-type galaxies is probably infall of star clusters through dynamical friction, while at higher masses, gas accretion resulting from mergers and torques becomes dominant. 

\begin{figure}
\resizebox{\hsize}{!}{\includegraphics[angle=0]{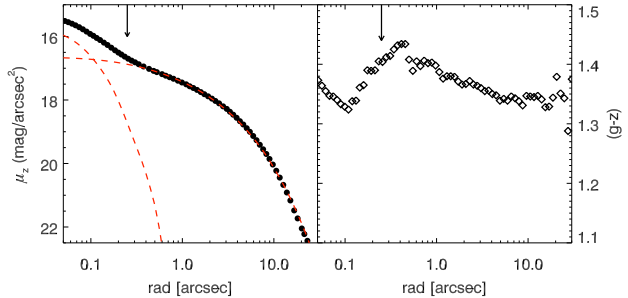}}
\caption{\label{fig:acs_colour} An HST/ACS $z$-band profile {\it (left panel)} and $(g-z)$ colour {\it (right)} of FCC\,277 \citep{turner12,ferrarese13}. The coloured curves show the S\'{e}rsic fits to the two components: nucleus and galaxy. The arrows at 0\farcs25 indicate the break radius, \rb, at which the nucleus component starts to dominate the galaxy light.
}
\end{figure}

In this paper we present a detailed study of the nucleus in the  intermediate-mass early-type galaxy FCC\,277 (NGC\,1428) and  discuss our observations  in the light of the current assumptions of nucleus formation. This galaxy is a member of the Fornax cluster and is part of the ACS Fornax Cluster Survey \citep{jordan07}. Its basic properties, as well as the main parameter of the NSC, are listed in Table~\ref{tab:fcc277_tab}.  In Fig.~\ref{fig:acs_mge} we show part of the HST/ACS $z$-band image, together with the field-of-view of VLT/SINFONI that we used to complete our study. In Fig.~\ref{fig:acs_colour} {\it (left panel)} we plotted the $z$-band surface brightness profile, together with the two S\'{e}rsic fits (dashed lines) that describe the galaxy light. The outer galaxy light is represented with a S\'{e}rsic fit with $n=1.8$. The point where the nucleus starts to dominate over the inner extrapolation of the S\'{e}rsic fit is called break radius and for FCC\,277 has the value \rb$=0\farcs25$ (indicated with an arrow in Fig.~\ref{fig:acs_colour}). The nucleus is fitted with another S\'{e}rsic profile with $n=1.7$. For a detailed description of the fitting process see \citet{turner12,ferrarese13}.

%

 The paper is organised as follows:  in Sect.~\ref{sec:obs_datared} we describe our observations and data reduction. In Sect.~\ref{sec:comp_sinf_hst} we compare the light profiles obtained from HST/ACS and VLT/SINFONI coupled with adaptive optics. Sect.~\ref{sec:kinematics} is devoted to the kinematics analysis of our IFU data, while in Sect.~\ref{sec:dyn_model} we present results from dynamical modelling of the galaxy and the NSC. In Sect.~\ref{sec:stellar_pop} we explore the stellar populations of the nucleus of this galaxy. In Sect.~\ref{sec:discussion} we discuss our findings in the light of  current galaxy and NSC formation models. We conclude in Sect.~\ref{sec:conclusions}.

%
%
\begin{table}
\caption{\label{tab:fcc277_tab} Basic properties of FCC\,277 and its Nuclear Star Cluster.}
\flushleft
\begin{tabular}{l l c }
\hline 
\hline 
FCC\,277 & & Reference \\
\hline
Morphological Type & E5 & \citet{ferguson89}\\
B$_{T}$ & 13.8$^{m}$\ & \arcsec \\
Effective radius & 10.2\arcsec & \arcsec \\
$(g-z)$ colour & 1.31$\pm$0.01 & \citet{blakeslee09}\\ 
Distance & 20.7$\pm$0.7 Mpc & \arcsec\\
Major axis position angle & 115\degr & \citet{graham98} \\
Velocity dispersion & 81.7 \kms & \citet{wegner03}\\
$M_\mathrm{vir}$& $ \sim 8\times10^{9} M_{\odot}$ & $^{1)}$\\
\hline
\hline
NSC & & \\
\hline
Effective radius & 0.09\arcsec$\sim$ 9 pc & \citet{turner12}\\
$g$ (mag) & 20.08$^{m}\pm$0.16 & \arcsec \\
$(g-z)$ & 1.33$\pm$0.18& \arcsec\\
\hline
\hline
\end{tabular}
{$^{1)}$ Using $M_{vir} =  5.0 \, R_\mathrm{eff} \, \sigma^{2}/G$ \citep{cappellari06} }
\end{table}

\section{Observations and data reduction}
\label{sec:obs_datared}

\subsection{Observations}
\label{sec:obs}

We obtained  integral field spectroscopy of the Fornax E5 galaxy FCC\,277 (NGC\,1428) using VLT/SINFONI \citep{eis03,bonnet04} in Natural Guide Star adaptive optics mode on  October 6, 7, and 10, 2007  (programme ID 380.B-0530, PI L. Infante). We used the $K$-band grating (1.95 -- 2.45 $\mu$m) that gives a spectral resolution R$\sim$3500 (6.2 \AA\/ FWHM as measured on sky lines). Our observations cover the central 3\arcsec $\times$ 3\arcsec, with a spatial sampling of 0\farcs05 $\times$ 0\farcs10. As a natural guide star we used a $R=14^{m}$  star located at 3\farcs5 to the North of the galaxy centre  (see Fig.~\ref{fig:acs_mge}). Due to its proximity to the galaxy centre, this star does not appear in the guide star catalogue or the USNO catalogue as a separate entry. Its celestial coordinates are $\alpha(J2000) =$ 03:42:22.9 and $\delta(J2000) =$ -35:09:11.4.  Our observations were carried out in service mode.

For the observations we used the standard near-IR nodding technique. Each observing block consisted of a sequence of object and sky frames (OOSOOSOOS), each individual integration was 300 s,  the sky fields were offset by 50\arcsec to the North. Science frames were dithered by 0\farcs05 and 0\farcs15 in order to reject bad pixels. There were six observing blocks.  The total on-source integration time was 3 hours. Additionally, after each observing block and at a similar airmass, we observed a B dwarf to act as a telluric star.

%
\begin{figure}
\resizebox{\hsize}{!}{\includegraphics[angle=0]{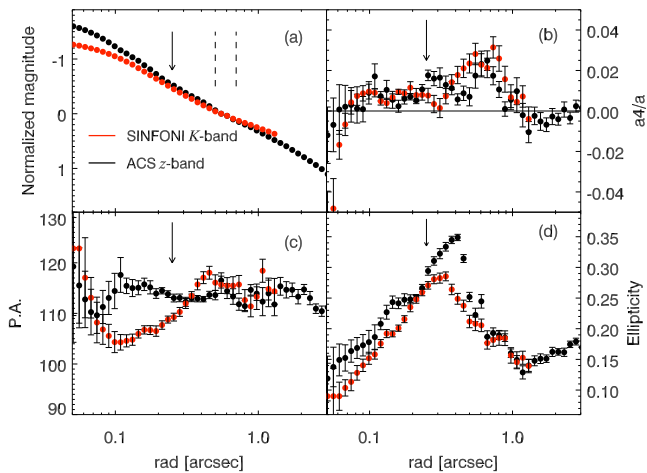}}
\caption{\label{fig:light_profile} Comparison between isophotal parameters obtained from space and  adaptive optics ground-based imaging. The red symbols show the light profile of FCC\,277 as derived from our SINFONI $K$-band reconstructed image. The black symbols denote $HST/ACS$ $z$-band imaging. The two vertical dashed lines in panel {\it (a)} show the area where the two profiles were normalised (0\farcs5 $<r<$ 0\farcs7). The vertical arrows in each panel indicate the break radius, as in Fig.~\ref{fig:acs_colour}.  In panel {\it (b)} we compare the $4^{th}$ cosine term of the isophotal fits that is indicative for deviations from pure elliptical shape. In panel {\it (c)} the measured position angles are displayed, and in panel {\it (d)} the ellipticity of the isophotes is shown.
}
\end{figure}

\subsection{Data reduction}
\label{sec:data_red}

We used the ESO SINFONI pipeline v2.0.5 to perform the basic data reduction on each observing block, consisting of six object and three sky exposures. In brief, the pipeline extracts the raw data, applies distortion, bad pixels and flat-field corrections, wavelength calibration, and stores the combined sky-subtracted spectra from one observing block in a 3-dimensional data cube. For each resulting data cube, we then ran the {\tt lac3d} code  \citep{davies10} to detect and correct residual bad pixels identified using a 3D Laplacian edge detection method.

We reduced the telluric stars in the same way as the science frames. Then for each telluric star we extracted a one-dimensional spectrum, removed the hydrogen Brackett\,$\gamma$ absorption line at $2.166\,\mu$m after fitting it with a Lorentzian profile, and divided the star spectrum by a black body spectrum with the same temperature as the star. The last step in preparing the telluric spectrum was to apply small shifts ($<$0.05 pixels) and scalings to minimise the residuals of the telluric features. To do this, we extracted a central one-dimensional spectrum from each science data cube and cross-correlated and fitted it with the corresponding telluric spectrum. Then we divided each individual spaxel in the six galaxy data cubes by the corresponding  best fitting telluric spectrum. In this way we also obtained a relative flux calibration.

Finally, we combined the six galaxy data cubes, using a $3\sigma$-clipping pixel reject algorithm. We also reconstructed a two-dimensional image of the galaxy, after integrating the spectral dimension of the final data cube in the range 2.1 -- 2.4~$\mu$m, where the contamination from sky lines residuals is minimal. To be able to robustly measure the velocity and velocity dispersion from the spectra of the galaxy a minimum signal to noise of 20 per pixel is required. Because the SINFONI pipeline does not provide error propagation during data reduction, we estimated the noise in each spectrum of the data cube as the r.m.s. of the residuals after subtracting a smoothed model of the spectrum. Then we used this noise estimate to bin the final galaxy data cube to achieve an approximately constant S/N$\sim$25 using the Voronoi 2D binning method of \citet{cc03}. This S/N allowed us to conserve a good spatial resolution (our smallest bins in the centre are $\sim$0\farcs1 across), while we are still able to reliably extract the stellar kinematics.

\section{Comparison between VLT/SINFONI and HST/ACS light profiles}
\label{sec:comp_sinf_hst}

%
\begin{figure}
\resizebox{\hsize}{!}{\includegraphics[angle=90]{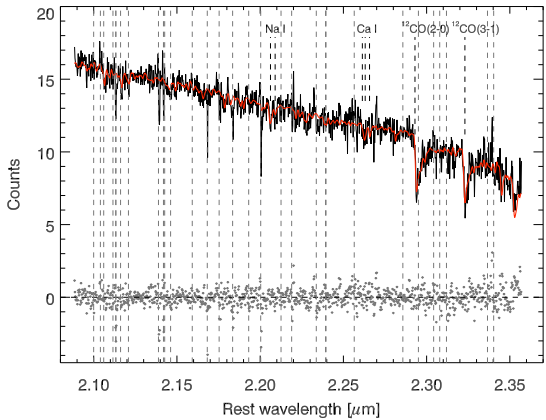}}
\caption{\label{fig:fit_spec} The spectrum from the central bin (S/N$\sim$25) of FCC\,277 with an over-plotted best fitting template spectrum, derived by the pPXF code  (red line; fit residuals are shown in grey). The location of the strongest near-IR absorption features in the $K$-band is indicated. The dashed vertical lines show the location of the strongest sky emission lines.
}
\end{figure}

In Fig.~\ref{fig:light_profile} we compare the light profiles of FCC\,277 derived from HST/ACS imaging and VLT/SINFONI observations. This galaxy is part of the ACS Fornax Cluster Survey \citep{jordan07} and high spatial resolution imaging in the $g$-band and $z$-band from the ACS is available. We used the IRAF task {\em ellipse} to fit elliptical isophotes to the ACS $z$-band image (black symbols in Fig.~\ref{fig:light_profile}) and to the SINFONI reconstructed image (before binning; red symbols). In panel {\em (a)} the two luminosity profiles are compared, after being normalised in the region 0\farcs5 -- 0\farcs7 (dashed vertical lines). Isophotal parameters are plotted against  the semi-major axis length. 
 The vertical arrows denote the break radius, \rb$=0\farcs25$, at which point the nuclear component starts to dominate the surface brightness profile \citep[][see also Fig.~\ref{fig:acs_colour}]{cote07}.

Although observed with adaptive optics, the SINFONI light profile is less steep  than the ACS profile within the inner $\sim$0\farcs3. Assuming this is due to the lower spatial resolution of the SINFONI data, we estimated  their PSF to be 0\farcs165 (FWHM) by convolving the ACS image \citep[using a Tiny Tim PSF,][]{krist95} with a given  Gaussian PSF until it matches the light distribution in the SINFONI image  (observations of stellar PSFs were not obtained during the SINFONI run). 

In panel  {\em (b)} of Fig.~\ref{fig:light_profile} we plotted the cosine $4^{th}$ order Fourier coefficient of the isophotes, divided by the semi-major axis length. Positive values of this parameter are indicative of disky isophotes,  as are observed in both the $z$- and $K$-band images in the inner 1\arcsec. There are two peaks in the $a4/a$  profile, one at $\sim$0\farcs2, coinciding with the break radius, and a stronger second peak at $\sim$0\farcs6, coinciding with the peak in the velocity field (see Sect.~\ref{sec:kinematics}). In panel {\em (c)} no significant variations of the position angle outside of the break radius are observed and the mean PA is consistent with the one derived at larger radii (see Table~\ref{tab:fcc277_tab}). In panel  {\em (d)} the ellipticity  reaches a maximum at $\sim$0\farcs35 for both the ACS and SINFONI profiles, although with different amplitudes. These differences are expected due to the differences in the PSF of the two images; a larger PSF leads to rounder isophotes \citep{peletier90}.  The comparison of the two profiles led us to the conclusion that the SINFONI ground based adaptive optics assisted observations are similar in quality to the HST/ACS images.

%
\begin{figure}
\resizebox{\hsize}{!}{\includegraphics[angle=0]{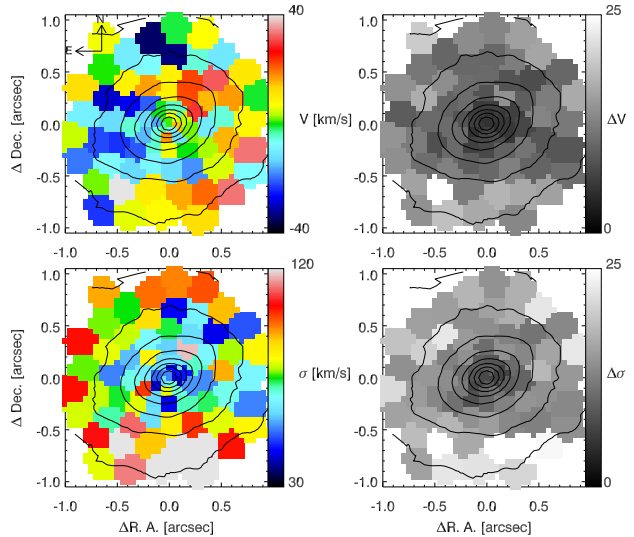}}
\caption{\label{fig:kin_maps} Velocity ({\it top panel}) and velocity dispersion ({\it bottom panel}) maps of FCC\,277 and corresponding errors {\em (right panels)}. Over-plotted are contours with constant surface brightness, as derived from our reconstructed SINFONI $K$-band image.}
\end{figure}

The observed features of the isophotal parameters point to a picture where, within the break radius,  the nuclear star cluster may be flattened  or, alternatively, may be the superposition of a round NSC and a larger scale disk. Such nuclear disk  beyond the break radius is evident in the diskiness parameter $a4/a$ at $\sim$0\farcs6 \citep[see also][]{turner12}.

\section{Stellar kinematics}
\label{sec:kinematics}

%
\begin{figure}
\resizebox{\hsize}{!}{\includegraphics[angle=0]{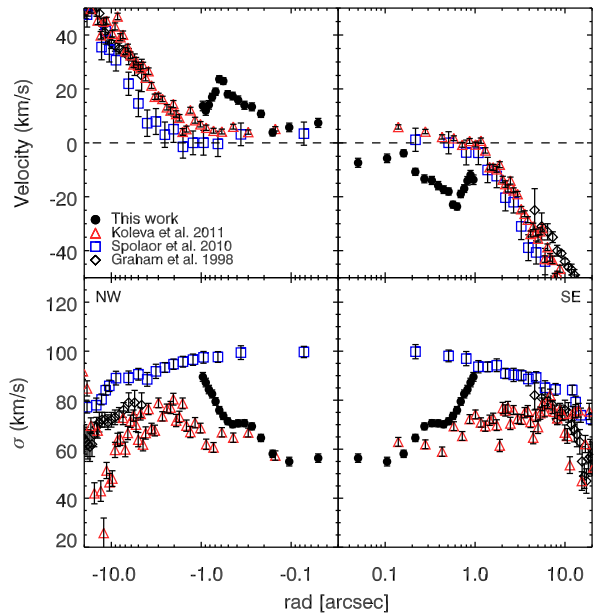}}
\caption{\label{fig:rad_kin} Behaviour of the stellar kinematics at different spatial scales.  Long-slit kinematics data from \citet[][diamond symbols]{graham98}, \citet[][blue squares]{spolaor10} and \citet[red triangles]{koleva11} are shown.   With solid symbols we plotted the innermost kinematics profile of FCC\,277, extracted from the SINFONI maps using kinemetry.}
\end{figure}

We used the pPXF code \citep{ce04} to derive the first and second order of the line-of-sight velocity distribution, working with a library of seven template spectra of K and M giant stars. These templates were observed with the same instrument and the same setup as our science target. To find the best fitting  composite template spectrum we used the region between 2.1 and 2.36~$\mu$m, where several strong absorption features allow accurate measurements (see Fig.~\ref{fig:fit_spec}), and masked the strong near-IR sky lines \citep{rousselot00}. In Fig.~\ref{fig:fit_spec} we show the spectrum of a central bin in the galaxy with the over-plotted best fitting composite template as derived by the pPXF code (in red), as well as the residuals (in grey).

Our stellar mean velocity and velocity dispersion maps  are shown in Fig.~\ref{fig:kin_maps}. Using kinemetry, described by \citet{krajnovic06}, we extracted the velocity and velocity dispersion profiles, shown with filled symbols in Fig.~\ref{fig:rad_kin}.  We observe rotation around the minor axis of the galaxy up to $\pm$25~\kms at $r\sim$0\farcs6, which is outside the break radius of the luminosity profile (\rb$=0\farcs25$, see Fig.~\ref{fig:acs_colour}, left panel) where the NSC is supposed to reside. At the centre of the galaxy the velocity dispersion approaches $\sim$55~\kms and then it increases in the outer parts of the field of view to reach $\sim$90 \kms at $r\sim$~1\arcsec. 

In Fig.~\ref{fig:rad_kin} we also compare our own SINFONI data with a compilation of literature  measurements, derived using long slits, aligned along the major axis of the galaxy. \citet[diamond symbols]{graham98} used a spectrograph with a slit width 2\arcsec\/ on the Australian National University’s 2.3 m telescope at Siding Spring Observatory. Their data do not cover the inner 4\arcsec of the galaxy, due to the presence of a relatively bright star close to the nucleus (that we used as a natural guide star for the AO). \citet[blue squares]{spolaor10} used the GEMINI/GMOS instrument with a slit width of 1\arcsec. The seeing during these observations was in the range 0\farcs7\,--\,1\arcsec. \citet[red triangles]{koleva11} reanalysed the same observations.

Based on Fig.~\ref{fig:rad_kin} we conclude that the rotating substructure in the nucleus of FCC\,277 co-rotates with the main body of the galaxy and that the velocity dispersion in the outer regions of the SINFONI field-of-view reaches similar values as the long slit studies. We note that the spatial resolution of our adaptive optics supported data is much higher than the resolution achieved by the other three studies, thus we cannot directly compare the radial profiles at small galactocentric radii. The significantly worse spatial resolution of the long-slit observations means that the inner rotation and velocity dispersion dip are washed out.

The observed rotation, taken together with the drop in the velocity dispersion, indicates the presence of a co-rotating cold substructure in the inner 0\farcs6\/ of the galaxy. We fitted the kinematic position angle of this substructure using the method described in Appendix C of \citet{krajnovic06}. The measured value is 118\degr$\pm10\degr$, which is consistent with the photometric position angle of the main body of the galaxy, derived at larger radii (see Table~\ref{tab:fcc277_tab}). 

For early-type galaxies the apparent stellar angular momentum $\lambda_{R_e}$ and the galaxy flattening are now a commonly used tool to classify galaxies into \emph{fast}- and \emph{slow}-rotators \citep{emsellem07,cap07}. The method needs IFU data to measure $\lambda_{R_e}$ inside one-effective radius. For FCC\,277 only long slit data are available out to the effective radius. Thus we used our best-fit Schwarzschild model  from Sect.~\ref{sec:dyn_model} to simulate the velocity and velocity dispersion as they would be observed by an IFU. We measured $\lambda_{R_e}=0.3$. Using the most recent classification from \citet{emsellem11}, we found that this galaxy is a \emph{fast}-rotator and lies slightly above the dividing line between the two classes, with its ellipticity $\epsilon=0.3$.

%
\begin{figure*}
\resizebox{\hsize}{!}{\includegraphics[angle=0]{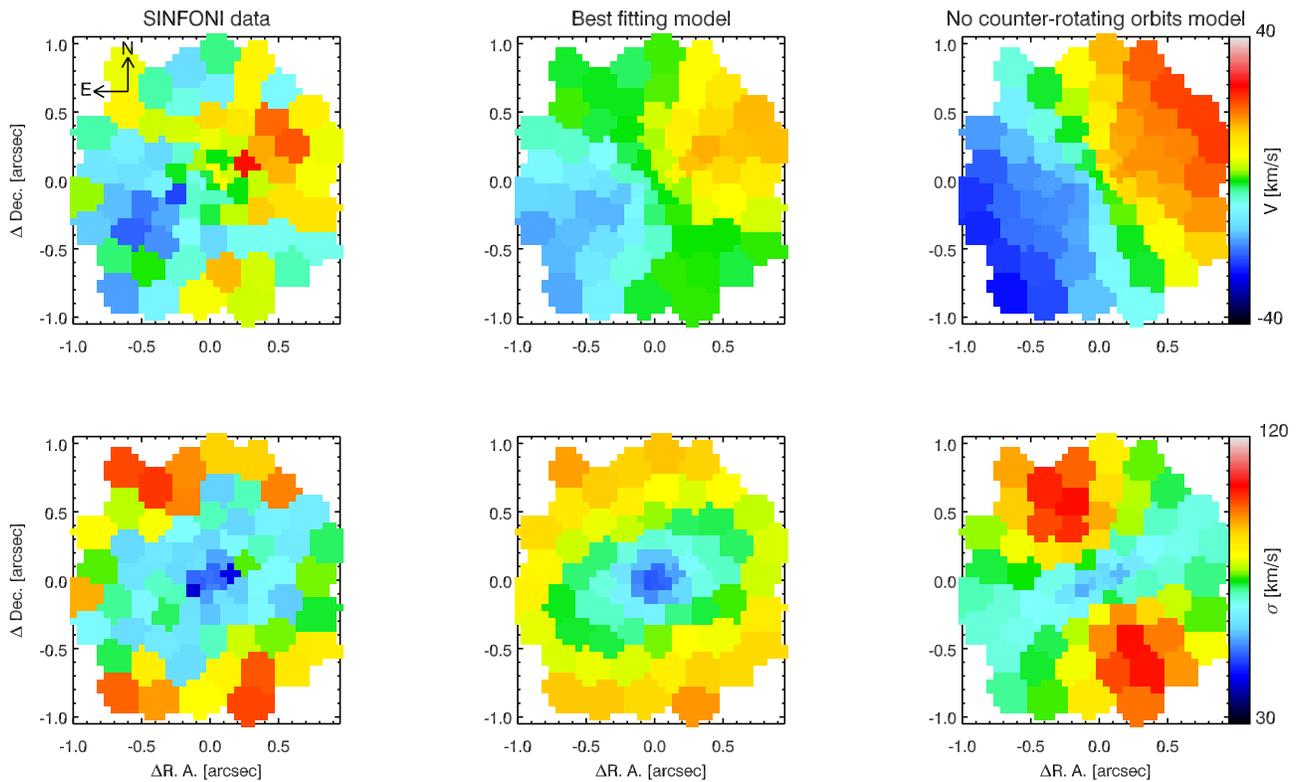}}
\caption{\label{fig:dyn_model_data} Comparison between our symmetrised SINFONI kinematics {\em (left panels)} and the ones obtained by the best fitting dynamical model {\em (middle panels)}.  The {\em right panels} show the resultant kinematics maps from our best fitting model when we do not include stars on counter-rotating orbits.
}
\end{figure*}

\section{Dynamical modelling}
\label{sec:dyn_model}

%
\begin{figure}
\resizebox{\hsize}{!}{\includegraphics[angle=0]{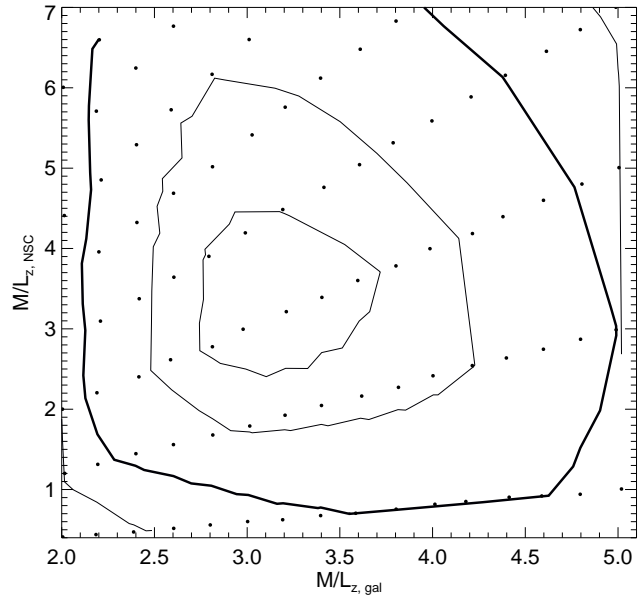}}
\caption{\label{fig:MLSCcontours} Confidence interval of the dynamical models of FCC\,277 for the mass-to-light ratio in $z$-band. The black dots indicate the location of the models and the contours indicate 1, 2, and 3$\sigma$ intervals, where the 3-sigma level is indicated by a thick line.
}
\end{figure}

To measure the mass distribution and the orbit configuration of the  inner part of FCC\,277 we used \citet{schw79} modelling. This method \citep{rvdb08} works by constructing a trial mass model of the galaxy, including a black hole, stars and dark halo. Then, the gravitational potential is inferred from the mass model and representative orbits are integrated numerically, while keeping track of the paths and orbital velocities of each orbit. We can then create a mass model of the galaxy by assigning an amount of mass to each orbit so that the overall stellar mass distribution is reproduced, while simultaneously fitting the observed stellar kinematics. The effect of the PSF on the observed stellar kinematics is an integral part of the dynamical model. These models have the advantage that they do not require any assumptions regarding the orbital anisotropy of the galaxy.

The models were constructed as follows. First, we parametrized the galaxy stellar surface brightness using the multi-Gaussian expansion (MGE) method, described by \citet{cap02}, on the ACS $z$-band image. In Fig.~\ref{fig:acs_mge} we show this image with overlaid contours of the MGE light model. There were 13 Gaussians with varying flattening fitted in total, the first two of them describing the NSC. The galaxy shows strong rotation around the minor axis and we therefore assumed the galaxy is oblate axisymmetric, which is the most common configuration \citep[e.g.][]{padila08}. The galaxy is also strongly flattened, with a minimum flattening 0.6 at 15\arcsec, and can thus not be seen more face-on than $i=65^{\circ}$. Then we used our  symmetrised  \citep[using the method described in Appendix A of][]{rvdb10} SINFONI IFU kinematics from Sect.~\ref{sec:kinematics} and the Schwarzschild orbit superposition method \citep{rvdb08} to construct a realistic dynamical model for the galaxy and the NSC. We also included the long-slit data of \citet{graham98}, to be able to constrain the mass-to-light ratio of the main body of the galaxy.  We did not include the data of \citet{spolaor10} and \citet{koleva11}, which result from two different reductions of the same data set, because they do not match with our kinematics measurements for the inner parts of the galaxy.  We probed the following parameters: the central black hole mass and separate mass-to-light ratio (M/L) for the galaxy and the NSC, using 5000 dynamical models. Changes in the inclination between 65\degr and 90\degr\/  led to insignificant changes in the M/L, thus we marginalised over it.

In Fig.~\ref{fig:dyn_model_data} we show the input symmetrised SINFONI kinematics velocity and velocity dispersion maps {\em (left panels)} together with the resulting kinematics obtained by the best fitting dynamical model {\em (middle panels)}.  The reduced $\chi^2$ of the best models is $\sim$0.21 over the 81 SINFONI bins.  The low value of the reduced $\chi^2$ is due to the very conservative estimate of our kinematics errors. The best-fit $M/L_{z}$ of the galaxy and NSC is $3.2\pm0.4$ and $3.0\pm1.0$ respectively, as shown in Fig.~\ref{fig:MLSCcontours}. Confidence intervals are determined using $\Delta\chi^2$ statistics, assuming  two degrees-of-freedom. Thus, the mass of the NSC is $1.4\pm 0.4 \times 10^{7} M_{\odot}$. The black hole mass is unconstrained, as the uncertainties on the central kinematics are too large. The difference in velocity dispersion between a 10$^{5}$  and a 10$^{7} M_{\odot}$ black hole is 5\,\kms\/ and the uncertainties on $\sigma$ are $\sim15$\,\kms.   Black hole masses above 10$^{7} M_{\odot}$ do yield significantly worse fits and hence we place an upper limit of 10$^{7} M_{\odot}$. To robustly determine the black hole mass higher S/N spectra of the nucleus need to be obtained to reduce the uncertainties on the central kinematics and the PSF of the IFU data needs to be known precisely. The inclusion of a dark matter halo does not alter the $M/L$ of the NSC and only very weakly the $M/L$ of the galaxy. The $M/L_\mathrm{gal}$ is expected to contain only a small contribution from the dark matter \citep{cappellari06}, hence our final adopted model does not include a dark matter halo. Kinematics reaching much further out are need to properly constrain the dark matter halo.

\begin{figure}
\resizebox{\hsize}{!}{\includegraphics[angle=0]{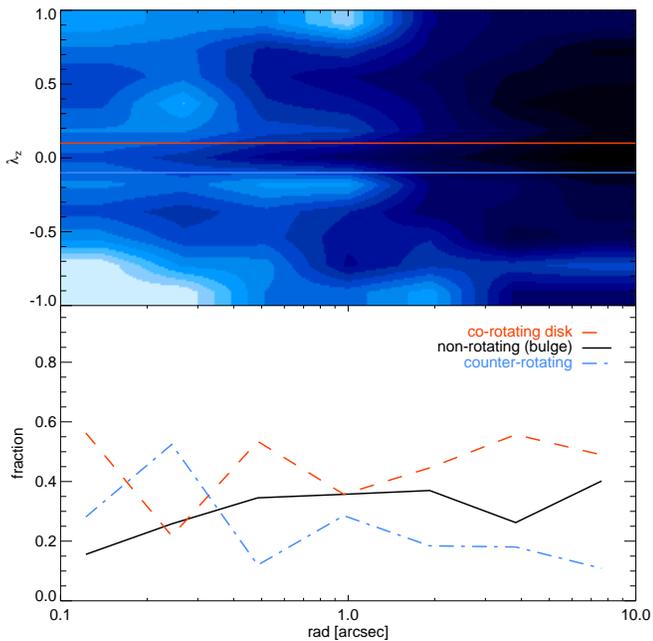}}
\caption{\label{fig:frac_orbits}  {\em Top panel:} Distribution of mass along the orbits of our best fitting dynamical model, as a function of angular momentum and radius. Colour coding reflects a factor of 3.5 span in mass density (darker colour corresponds to higher mass).
 {\em Bottom panel:} Fraction of non-/rotating orbits as a function of radius. A non rotating bulge ($-0.1<\lambda_{z}<0.1$) is denoted with a black line. The red line corresponds to the sum of all  co-rotating orbits with $\lambda_{z}>0.1$, the blue line - all  counter-rotating orbits with  $\lambda_{z}<-0.1$}
\end{figure}

Apart from the mass distribution of the galaxy, the models  yield the orbital distribution as a function of radius. In Fig.~\ref{fig:frac_orbits} we show the  orbital mass weights as a function of the average radius and spin $\bar \lambda_z = \bar J_z \times (\bar r / \bar \sigma)$, where $\bar J_z$ is the average angular momentum along the short $z$-axis and $\bar \sigma$ the average second moment of the orbits. 

We detect the presence of three distinct components: both a  co- and  counter-rotating component as well as a non-rotating bulge component. The relative contribution of each of these components is shown in the {\em bottom panel} of Fig.~\ref{fig:frac_orbits}. Both rotating components extend well inside the break radius and have similar contributions in the NSC region. The sigma drop seen in the stellar dispersion map coincides with  a decrease of the non-rotating orbits. The question arises if this is the only possible orbital configuration for this system. The orbits available to the Schwarzschild models are a fully representative set and the linear solver used to construct the models is guaranteed to find the global minimum \citep{vdv08,rvdb08}. This guarantees that the model finds the best-fitting orbital configuration. There could be other solutions that are also a good representation of the observations.  As a consistency check, we attempted to fit models without  counter-rotating orbits,  which would exclude an opposite angular momentum accretion event as a formation scenario for the nucleus. However this led to a significantly worse match of the stellar kinematics  (see Fig.~\ref{fig:dyn_model_data}, {\em right panels}),  which indicates that counter-rotating orbits are thus required.  

We note that we were unable to use the Jeans' modelling approach to fit the stellar kinematics using the method described by \citet{cappellari08}. Although we could receive  reasonable fits to the second velocity moment ($V_\mathrm{RMS}$), the fits to the velocity field were unsatisfactory. This is because, as it is at the moment, the JAM package does not allow the rotation parameter $\kappa$ to accept positive and negative values simultaneously for a given MGE Gaussian.

To be able to quantitatively discuss the different orbital fractions, the dark matter halo of the galaxy and its global M/L, one would need improved  long-slit or other large scale kinematics.

%

\section{Stellar population parameters}
\label{sec:stellar_pop}

 In addition to the structure and dynamics, we can investigate what the  parameters of the stellar populations of the NSC and/or nuclear disk in the heart of FCC\,277 are and whether they differ from the main body of the galaxy. 

Usually, nuclei in low luminosity Fornax and Virgo galaxies are bluer compared to their hosts \citep{cote06,turner12}. In the right panel of Fig.~\ref{fig:acs_colour} we show the $(g-z)$ colour profile, as derived from HST/ACS imaging \citep{turner12,ferrarese13}. The  integrated colour of the nucleus is $(g-z)=1.33\pm0.18$ \citep{turner12} and does not differ from the main body of the galaxy.  If nuclei follow the same colour-metallicity relation as globular clusters in early-type galaxies do \citep{peng06}, than the  red colour would be indicative for higher metallicity of the NSC. However, age effects  cannot be excluded, due to the well known age-metallicity degeneracy of broad band colours.

\citet{koleva11} measured  from optical spectroscopy the age and metallicity of the core of FCC\,277 (within a 0\farcs5 radius aperture) to be 5.4~Gyr and [Fe/H]=$-0.07$, respectively. At the effective radius these values are 7.7~Gyr and [Fe/H]=$-0.50$.  Their data lack the spatial resolution to differentiate the NSC and the disk,  however there is a pronounced negative age and positive metallicity gradient towards the nucleus.

We used our near-IR IFU spectra to measure the line strengths of \na\/ ($\sim$2.2~$\mu$m) and \co\/ ($\sim$2.3~$\mu$m) absorption features (see Fig.~\ref{fig:fit_spec}). From previous stellar population studies in the near-IR wavelength range we know that the \na\/ index increases with metallicity and younger age \citep{silva2008,esther09,az10a}, and the \dco\/ index is expected to increase with higher metallicity for ages above 3~Gyr \citep{maraston05}.

We measured the two indices using the definition of \citet{frog01} for the \na\/ index and  \citet{esther08} for the \dco\/ index.   Before measuring, we first broadened our spectra to 6.9~\AA\/ (FWHM, $\sim$~94~\kms) to match the spectral resolution of other stellar population studies of elliptical galaxies in the near-IR \citep[e.g.][]{silva2008,esther09}. Finally, we corrected the \na\/ index to zero velocity dispersions using the velocity dispersion corrections of \citet{silva2008}.

%
\begin{figure}
\resizebox{\hsize}{!}{\includegraphics[angle=0]{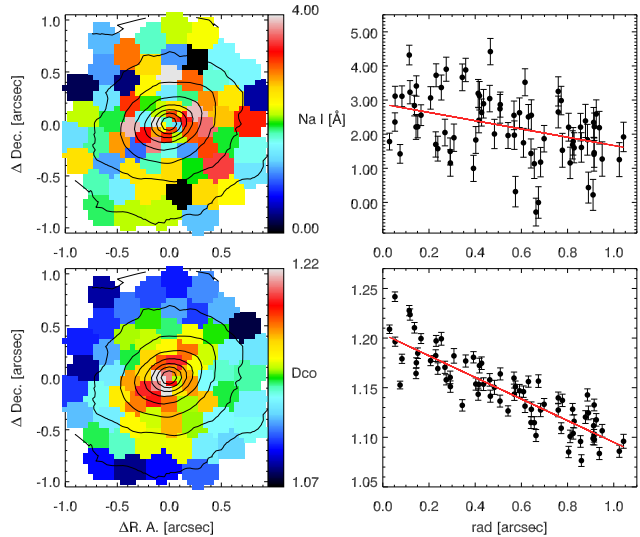}}
\caption{\label{fig:index_maps} \na\/ and \dco\/ index maps with HST like spatial resolution {\em (left panels)} and radial profiles {\em (right panels)}. The red solid lines are the least squares linear fits to the data.}
\end{figure}

In Fig.~\ref{fig:index_maps} we show our index maps as well as their radial profiles. We observe radial gradients for both indices. Increase in \na\/ towards the centre is consistent with increasing metallicity and/or younger age in the nuclear disk and NSC. On the \dco\/ map we see that the strongest index values towards the centre seem to form an elongated shape, aligned with the major axis of the galaxy and the rotating structure, visible on the velocity map (Fig.~\ref{fig:kin_maps}.)    This increase is again consistent with the redder colour and indicative of higher metallicity.

\begin{figure}
\resizebox{\hsize}{!}{\includegraphics[angle=0]{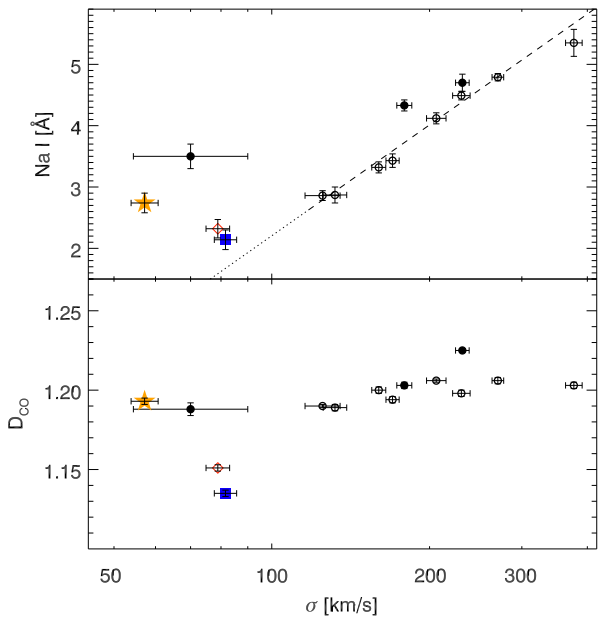}}
\caption{\label{fig:index_sig} \na\/ and \dco\/ indices versus the central velocity dispersion of Fornax early-type galaxies. Open symbols denote old Fornax galaxies, solid symbols - galaxies with optical signatures of recent ($<$3~Gyr) star formation \citep{silva2008}.  The dashed line represents the least-squares linear fit to the old Fornax galaxies only, the dotted line denotes its extrapolation to lower $\sigma$. The NSC in FCC\,277 is marked with an orange asterisk, the so called nuclear disk - with a red diamond, and the whole galaxy, as observed with SINFONI, is marked with a blue square. 
}
\end{figure}

 In Fig.~\ref{fig:index_sig} we  compare FCC\,277 with other early type galaxies in the Fornax cluster in terms of  their \na\/ and \dco\/ indices  versus their central velocity dispersion. The spectra of \citet{silva2008} cover 1/8 of the effective radius of each galaxy and are marked with circles. Open circles represent galaxies with old stellar populations, solid symbols stand for the galaxies that have optical signatures of recent ($<$3~Gyr) star formation.   The dashed line illustrates the least-squares linear fit to the old galaxies only. Our SINFONI observations of FCC\,277 cover approximately the same radial extent as the other galaxies. We plotted the values measured on the integrated spectrum with solid blue squares. We extracted the NSC area ($r\leq0\farcs25$) and marked our measurements with orange asterisks. With open red diamonds we indicated the values for the so called "nuclear disk", integrated over the range $0\farcs25<r\leq0\farcs7$.
 
The \na\/ index of the NSC is much higher compared to the  extrapolation (shown with a dotted line) of the $\sigma$-\na\/ relation for old Fornax galaxies and is closer to the systems with younger ($<$3~Gyr) stellar populations. There is not a big difference of the index strength between the nuclear disk and the galaxy as a whole, so one would infer purely old age for these two components if looking only at this plot.  The \dco\/ index seems to saturate for galaxies with velocity dispersion higher than 100~\kms\/. Until reliable stellar population models for the near-IR become available, we cannot provide a quantitative estimate for the changes in stellar population parameters in the nuclei of early-type galaxies. At the moment we can only speculate that, taken together with the red colour of the NSC (as red as the host galaxy, which is unusual for NSCs in Fornax as discussed above), these point to a mixture between younger age and higher metallicity compared to the main body of the galaxy.

%

\section{Formation mechanism of the NSC}
\label{sec:discussion}

So far we have collected evidence that: {\em i)} the nuclear star cluster  inside the central 0\farcs25 of FCC\,277 does not rotate  within our error of $\sim6$\,\kms; {\em ii)} the rotation visible at $\sim$0\farcs6 overlaps with a maximum in the diskiness of the isophotes, suggesting the existence of a nuclear disk around the NSC; {\em iii)} the existence of a central velocity dispersion drop is indicative  of significant  rotation in the same area; {\em iv)} dynamical modelling reveals a significant rotation in the inner 1\arcsec\/ in {\em both} directions, i.e.  co- and counter-rotation, explaining the low level of observed rotation, the low $\lambda_{R_e}$, and the sigma drop; {\em v)} there is no significant difference in the derived dynamical $M/L$ between the NSC and the galaxy. However, within the errors, a change in the stellar population parameters may not lead to an obvious change in the $M/L$ for the different components. The derived dynamical $M/L_{z}$ within the errors is consistent with the predictions of stellar population models for a Salpeter IMF and $5-10$~Gyr age \citep[e.g.][]{bc03,maraston05}.  If one compares to Chabrier or Krpupa IMF, then our dynamical $M/L_{z}$ is about a factor of two larger ; {\em vi)} there is evidence for differences in the stellar population parameters of the NSC, the nuclear disk, and the galaxy. This points to a scenario where the nucleus is younger and more metal rich.

All these evidences for  complex kinematics and stellar populations point to a scenario where the NSC and disk formed through multiple episodes of gas accretion and subsequent episodes of star formation.  Counter-roation points towards mergers with orbital angular momentum opposite to the host galaxy.  \citet{seth10} reached the same conclusion for the nucleus in NGC\,404, a nearby S0 galaxy. Thus, FCC\,277 is the second  early-type galaxy that has a NSC exhibiting  complex star formation history.  \citet{turner12} noted that the lowest-mass galaxies with nuclei in the ACS Fornax and Virgo Cluster Surveys seemed to be structurally simple, having likely formed through star cluster infall. The more massive galaxies (such as FCC\,277) seemed to be more structurally complex in their inner regions. They interpreted that as evidence for an increased importance of gas infall at higher masses. Our results about the nucleus of FCC\,277 are consistent with this picture. Certainly, a larger sample is needed to  study in more detail whether gas dissipation is a common mechanism for NSCs growth in early-type galaxies, as it is in late-type ones \citep[e.g.][]{walcher06,hartmann11}. The NSC accounts for $\sim0.2\%$ of the total mass of FCC\,277, thus it is not an outlier of the typical scaling relations in early-type galaxies \citep[e.g.][]{ferrarese06}.

\section{Concluding remarks}
\label{sec:conclusions}

In this paper we  present a pilot study of the detailed properties of nuclear star clusters (NSCs) in early-type galaxies.  Although the nucleation frequency in galaxies is estimated to be $\sim80\%$, detailed data about the chemical and dynamical properties of NSCs exist mainly for spiral hosts. In early-type galaxies this task observationally is not trivial due to the intrinsic brightness of the underlying galaxy light, as well as  small angular extent of the NSCs, which at the distance of Fornax, span $\sim$0\farcs1 in diameter. We showed that using current technology, it is indeed possible and valuable information about the formation mechanisms of NSCs can be obtained. As a pilot test case we chose the galaxy FCC\,277, a nucleated early-type galaxy that belongs to the Fornax cluster. This is the only   member of this galaxy cluster that has a conveniently located bright star that one can use as a natural guide star for the adaptive optics (AO) system. 

Using SINFONI AO assisted observations, we observed the central 3\arcsec$\times$3\arcsec\/ of the galaxy. Thus we obtained maps of the stellar kinematics with HST-like spatial resolution  of 0\farcs165 (FWHM). Our velocity map (Fig.~\ref{fig:kin_maps}) shows clear rotation with a maximum at $\sim$0\farcs6 from the galaxy centre, which overlaps with a maximum in the diskyness of the fitted isophots on the galaxy image (Fig.~\ref{fig:light_profile}). The NSC itself has an effective radius of 0\farcs08 and does not rotate  within our detection limit of $\sim6$\,\kms. However, we observe a pronounced drop in the velocity dispersion in the central 1\arcsec\/ that suggests the existence of a dynamically cold rotating sub-structure. Our dynamical modelling reveals that the nucleus of this galaxy is complex: co- and counter-rotating,  as well as non-rotating, stellar orbits are needed simultaneously to reproduce the observed kinematics (Fig.~\ref{fig:frac_orbits}). The NSC seems to be embedded in a disk that is most likely younger and more metal rich than the main body of the galaxy (Fig.~\ref{fig:index_sig}).  Due to insufficient S/N we can only provide a conservative upper limit for a possible black hole of 10$^{7} M_{\odot}$. All these facts  point to a complex formation history of the nuclear  region in FCC\,277. Most likely gas dissipation and merging played an important role in shaping the nucleus of this galaxy. To check whether this is a common phenomenon  among early-type galaxies, a larger sample is needed and can be obtained with current  observing facilities.


\section*{Acknowledgements}
  
We are grateful to the ESO astronomers who obtained the data presented in this paper in service mode operations at La Silla Paranal Observatory. We acknowledge fruitful discussions with Eric Peng, David R. Silva, Jakob Walcher, Hans-Walter Rix, Jesus Falc\'{o}n-Barroso. We thank Alister Graham, Max Spolaor, and Mina Koleva for providing us with their results in tabular form. ML would like to thank the staff at the Astronomical Observatory of the University of Sofia for their hospitality, where parts of this research have been carried out. LI and AJ acknowledge Fondecyt, Fondap and Basal funding for this project. AJ is supported by the Chilean Ministry for the Economy, Development, and Tourism's Programa Iniciativa Cient\'{i}fica Milenio through grant P07-021-F, awarded to The Milky Way Millennium Nucleus, by Anillo ACT-086 and BASAL CATA PFB-06.  We finally thank the referee for her/his valuable comments. This paper is dedicated to Mariika B. Ilieva (1926 - 2012) with a warm thank you for all the support.

\bibliographystyle{mn2e}
\bibliography{/Users/mlyubeno/Dropbox/sci/biblio/papers,/Users/mlyubeno/Dropbox/sci/biblio/9314}

\bsp 
\label{lastpage}

\end{document}